\documentclass[twocolumn]{IEEEtran}
\usepackage{amsopn,amsthm,amsmath,amssymb,amsfonts}
\usepackage{cite}
\usepackage{graphicx,subfig}

\newtheorem{example}{Example}
\newcommand{\eg}[1]{  \begin{small}
\begin{example} \hrulefill \\
 { #1 }
\end{example}
\end{small}
\hrulefill \\
}

\newcommand{\hl}{\hat{l}}

\title{Spatially Coupled Repeat-Accumulate Codes}
\author{Sarah Johnson and Gottfried Lechner
\thanks{The work of S. Johnson was supported by the Australian Research Council Grants DP0877258 and DP1093114. S. Johnson is with the University of Newcastle, Australia (e-mail: sarah.johnson@newcastle.edu.au.)

The work of G. Lechner was supported by the Australian Research Council Grant DP0881160. G. Lechner is with the Institute for Telecommunications Research at the University of South Australia (e-mail: gottfried.lechner@unisa.edu.au).}
}

\begin{document}

\maketitle

\begin{abstract}
In this paper we propose a new class of spatially coupled codes
based on repeat-accumulate protographs. We show that spatially
coupled repeat-accumulate codes have several advantages over
spatially coupled low-density parity-check codes including simpler
encoders and slightly higher code rates than spatially coupled
low-density parity-check codes with similar thresholds and decoding complexity (as
measured by the Tanner graph edge density).
\end{abstract}

\section{Introduction}

Convolutional LDPC codes, otherwise known as spatially coupled LDPC
codes (SC-LDPC), were first introduced by Felstr\"om and Zigangirov
in the late 90's \cite{Felstrom_IT99}. Performance results,
generated using either density evolution or decoding simulations,
have shown that SC-LDPC codes have excellent sum-product decoding
thresholds over a range of channels
\cite{Lentmaier_ISIT05,Tanner_IT04,Lentmaier_IT10}. Incredibly, and
in contrast to standard LDPC codes, these thresholds rapidly improve
as a function of the average Tanner graph node degree. This enables
the design of iterative error correction codes with both excellent
thresholds and very low error floors, something not so far achieved
with traditional LDPC or turbo codes.

Recent exciting developments have shown that the iterative decoding threshold of
certain SC-LDPC ensembles is actually equal to their MAP threshold on the binary erasure channel (BEC) \cite{Kudekar_IT11}. I.e., for spatially coupled codes iterative decoding is actually optimal on the BEC. It is conjectured, but not
yet proven, that this holds for more general channels as well.

In this paper we consider whether the concept of spatial coupling
can apply equally well to another class of iterative error
correction codes called repeat-accumulate codes.

Repeat-accumulate (RA) codes \cite{Divsalar_Turbolike}, are error
correction codes formed by the serial concatenation of a rate-$1/q$
repetition code and a $\frac{1}{1+D}$ convolutional code, called an
accumulator, with an interleaver, $\Pi$, and (optionally) a rate-$a$
combiner between them. Significantly, RA codes can be encoded using
serial concatenation of the constituent encoders, as for serially
concatenated turbo codes, and decoded using iterative decoding, as
for LDPC codes, thus gaining both the low encoding complexity of
turbo codes and the decoding performance of LDPC codes.

In this paper we will consider the spatial coupling of RA codes in
such a way to preserve the inherent advantage of RA codes, most
importantly their very simple encoding, while obtaining the
threshold advantages promised by the idea of spatial coupling.
Section~\ref{sec:SC-RA} introduces our proposed spatially coupled RA
codes, Section~\ref{sec:DE} presents threshold results derived using
density evolution and Section~\ref{sec:Sims} gives simulation
results comparing spatially coupled RA and LDPC codes.

\section{Spatially Coupled RA Codes} \label{sec:SC-RA}

Spatially coupled RA (SC-RA) codes can be formed in a similar manner
to spatially coupled LDPC (SC-LDPC) codes. We consider two ensembles,
the first we will use in practice to construct SC-RA codes,
and the second is useful to derive density evolution
equations.

\subsection{The (q,a,L) Ensemble}

\begin{figure}[t!]
    \begin{center}
        \includegraphics[width=0.9\columnwidth]{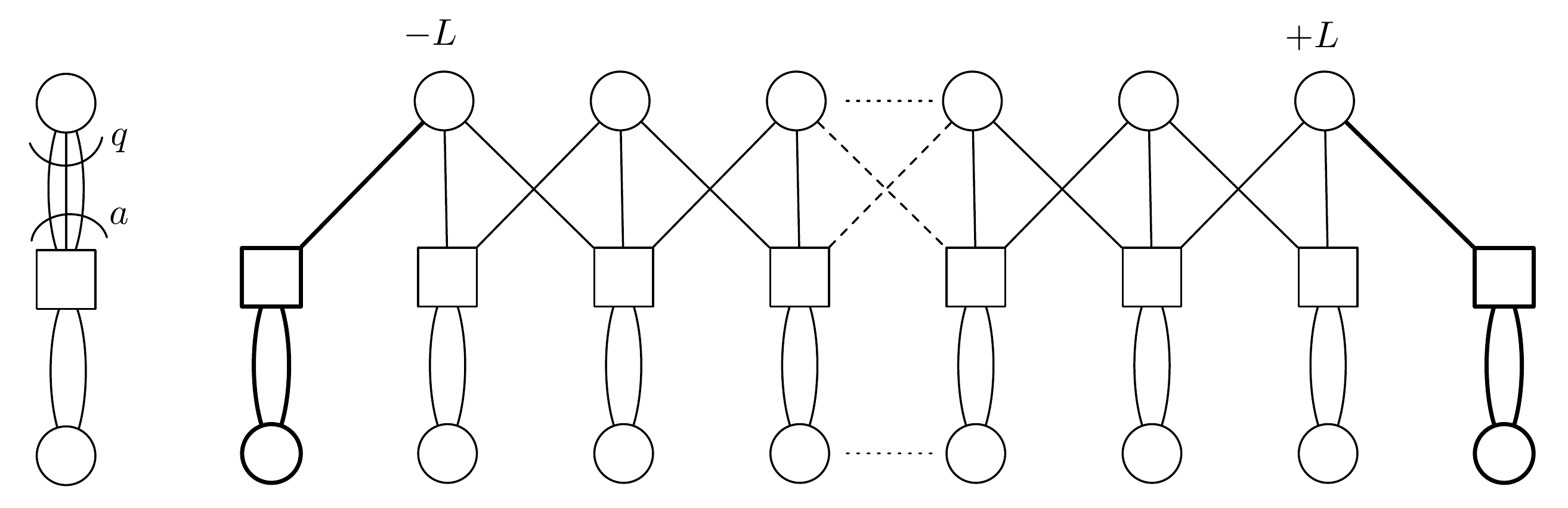}
        \caption{Coupled $q=3$, $a=3$ RA protographs.}
        \label{fig:GraphRA}
    \end{center}
\end{figure}

The left hand side of fig.~\ref{fig:GraphRA} shows the protograph of
a standard (3,3)-regular RA code. There is one message bit node,
shown at the top, a parity bit node, shown at the bottom, and a
check node in the middle. A coupled chain of $2L+1$ of these
protographs, shown on the right hand side of fig.~\ref{fig:GraphRA},
is formed by connecting each message bit to $\hl  = (q-1)/2$
protographs to the left and $\hl $ protographs to the
right\footnote{For the moment we assume $q$ is odd.}. As for coupled
LDPC chains we add $q-1$ extra check nodes (shown in bold) when
forming the coupled chain of protographs. For RA protographs we must
also add $q-1$ extra parity bit nodes (shown in bold) to avoid
creating any degree-1 check nodes.

We could have spatially coupled the parity bit nodes in the same way
as the message bit nodes, i.e. by connecting each parity bit node to
the check node of the protograph on the right hand side. However, if
the parity bit nodes are coupled in this way, the final code will
not retain the RA code accumulator structure. Keeping the parity bit
nodes uncoupled can be thought of as serially concatenating a
spatially coupled low-density generator matrix with a standard
accumulator.

A particular code from the $(q,a,L)$ ensemble will be formed using
copies of the coupled chain to give a total of $M$ message
bits per protograph. Our final code will thus consist of $(2L+1) M$ message
bit nodes, $(2L+1 + 2\hl )\frac{q}{a}M$ parity bit nodes and $(2L+1 +
2\hl )\frac{q}{a}M$ check nodes. Hence the code rate, assuming every
check node results in a linearly independent constraint, is
\begin{align}
    \label{equ:rateRA}
    r_{\text{RA}}   &= \frac{(2L+1)M }{(2L+1)M + (2L+1 + 2\hl )\frac{q}{a}M}\nonumber\\
                    &=  \frac{(2L+1)a }{(2L+1)a + (2L+q )q}.
\end{align}

When constructing a code from the $(q,a,L)$ ensemble each of the
message bit nodes at position $i \in \{-L, \cdots +L\}$ is connected
to exactly one of the check nodes at positions $j \in \{-i-\hl ,
\cdots i+\hl \}$. The choice of which of the check nodes to connect
to at each position can be chosen randomly.

For each protograph the $M$ parity-bit nodes are connected to the
$M$ check nodes in a traditional accumulator pattern. We also
connect the final bit node in each protograph to the first check
node in the following protograph. We could have separately
terminated the accumulator in each protograph (by connecting the last bit node in the protograph to the first bit node in the same protograph) to give $2L+1+2\hl$
separate size $M$ accumulators. For large enough $M$ there should not be a difference in performance, however, a single accumulator avoids the
$2L+1+2\hl$ length $2M$ cycles.

\begin{figure}[t!]
    \begin{center}
        \includegraphics[width=\columnwidth]{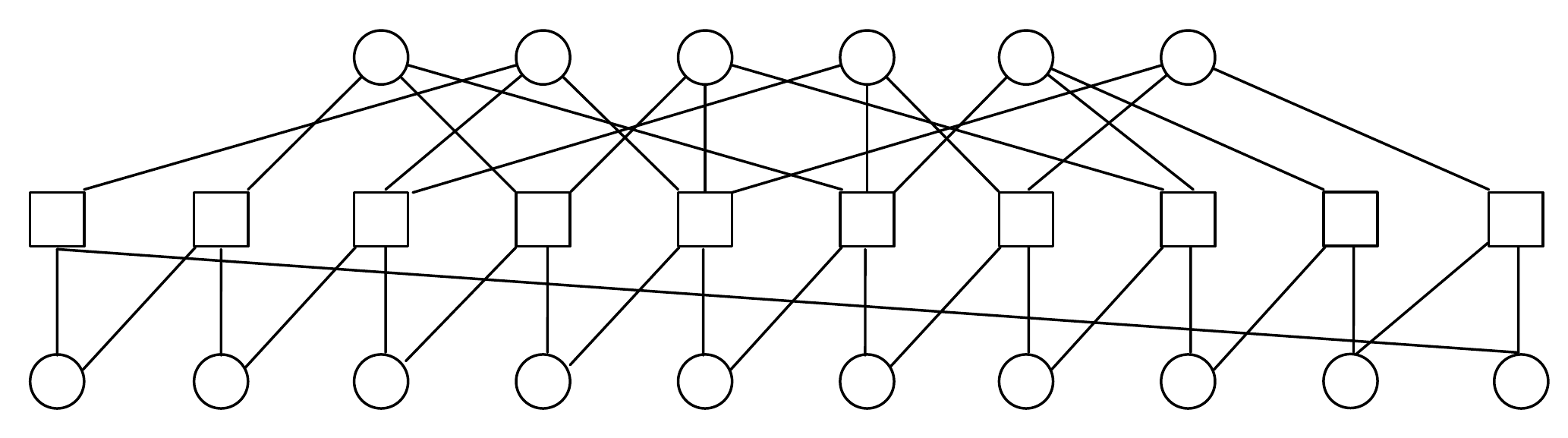}
        \caption{Example of a $q=3$, $a=3$ SC-RA code with $L=1$ and $M=2$.}
        \label{fig:GraphRA_eg5}
    \end{center}
\end{figure}

\eg{  A $(q=3,a=3)$ RA protograph is repeated $2L+1 = 3$ times to
give the coupled chain in fig.~\ref{fig:GraphRA}. Setting $M=2$ and
randomly choosing an edge permutation for the message bit edges
gives the SC-RA Tanner graph in fig. \ref{fig:GraphRA_eg5}. As SC-RA
codes are systematic we form the codeword using the messages bits
first, followed by the parity bits. This gives an SC-RA code with
parity-check matrix: $H = $
\[ \label{LDPC2_RA_H}  \left[
  \begin{array}{cccccccccccccccc}
     0 & 1 & 0 & 0 & 0 & 0 & 1 & 0 & 0 & 0 & 0 & 0 & 0 & 0 & 0 & 1 \\
     1 & 0 & 0 & 0 & 0 & 0 & 1 & 1 & 0 & 0 & 0 & 0 & 0 & 0 & 0 & 0\\
     0 & 1 & 0 & 1 & 0 & 0 & 0 & 1 & 1 & 0 & 0 & 0 & 0 & 0 & 0 & 0\\
     1 & 0 & 1 & 0 & 0 & 0 & 0 & 0 & 1 & 1 & 0 & 0 & 0 & 0 & 0 & 0\\
     0 & 1 & 1 & 0 & 0 & 1 & 0 & 0 & 0 & 1 & 1 & 0 & 0 & 0 & 0 & 0\\
     1 & 0 & 0 & 1 & 1 & 0 & 0 & 0 & 0 & 0 & 1 & 1 & 0 & 0 & 0 & 0\\
     0 & 0 & 0 & 1 & 0 & 1 & 0 & 0 & 0 & 0 & 0 & 1 & 1 & 0 & 0 & 0\\
     0 & 0 & 1 & 0 & 1 & 0 & 0 & 0 & 0 & 0 & 0 & 0 & 1 & 1 & 0 & 0\\
     0 & 0 & 0 & 0 & 1 & 0 & 0 & 0 & 0 & 0 & 0 & 0 & 0 & 1 & 1 & 0\\
     0 & 0 & 0 & 0 & 0 & 1 & 0 & 0 & 0 & 0 & 0 & 0 & 0 & 0 & 1 & 1 \\
 \end{array}
\right]
\]
In practice the edge corresponding to the Ô1Õ entry in the top right corner of $H$ is omitted for ease of encoding. \vspace{-1em}
}

\begin{figure}[t!]
    \begin{center}
        \includegraphics[width=0.9\columnwidth]{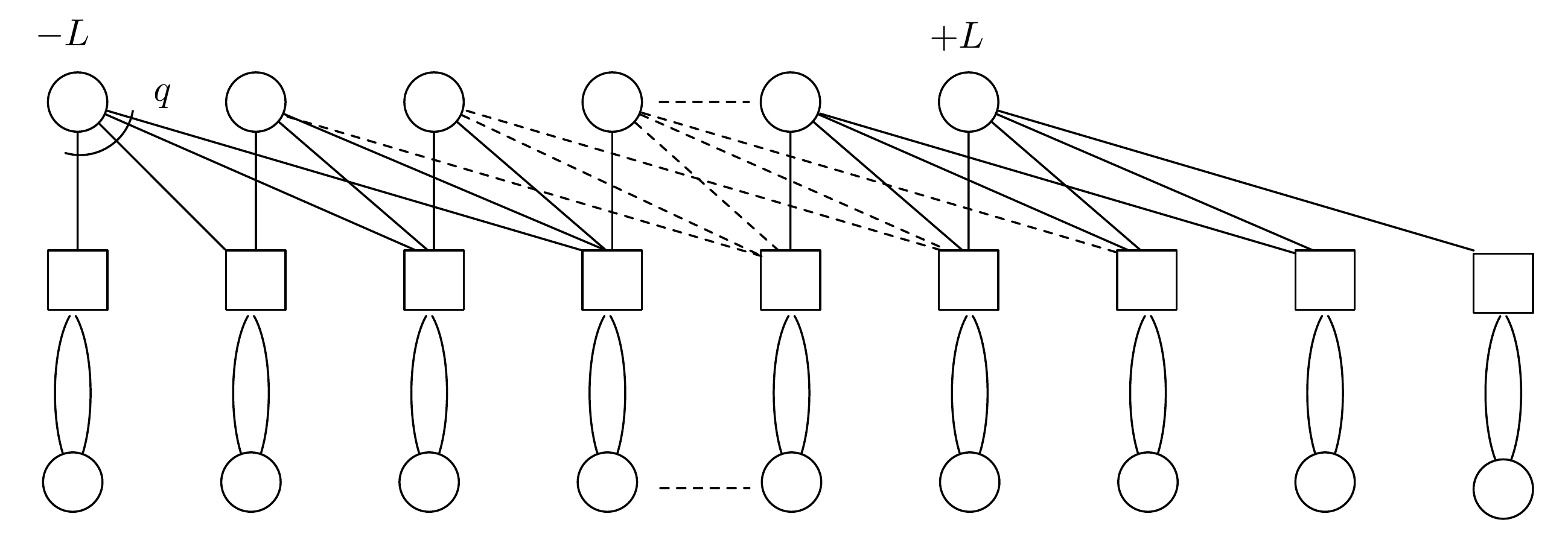}
        \caption{Coupled $q=4$, $a=4$ RA protographs.}
        \label{fig:GraphRAq4}
    \end{center}
\end{figure}

By slightly re-drawing fig.~\ref{fig:GraphRA} to push the top row of
nodes across to the left immediately shows how to construct SC-RA
codes with even values of $q$. Fig.~\ref{fig:GraphRAq4} for example
shows a SC-RA code with $q=4$.

\subsection{The $(q,a,L,w)$ Ensemble}

The ensemble $(q, a, L)$ can be modified by adding a ``smoothing"
parameter $w$ in a similar method to that for LDPC codes
\cite{Kudekar_IT11}. The $(q,a,L,w)$ ensemble is not used in
practice but is useful to simplify the derivation of density
evolution equations. Considering this ensemble for SC-RA codes will
allow a comparison of the asymptotic performance of SC-RA codes with
the SC-LDPC ensembles in \cite{Kudekar_IT11}.

As previously, at each position $[-L,L]$ there are $M$ message bit
nodes. However, the check nodes are considered to be located at all
integer positions $[- \infty, \infty]$ and there are $\frac{q}{a}M$
check nodes at each position. Only some of these positions actually
interact with the message bit nodes. Instead of requiring that each
message bit node at position $i \in \{-L, \cdots +L\}$ is connected
to exactly one of the check nodes at positions $j \in \{-i-\hl ,
\cdots i+\hl \}$ we assume that each of the $q$ connections of a
variable node at position $i$ is uniformly and independently chosen
from the range $[i, \cdots, i+w-1]$. Similarly, we assume that each
of the $a$ connections of a check node at position $i$ is
independently chosen from the range $[i-w+1,\cdots,i]$. $q$ need not
be odd. For simplicity we again assume that a parity bit node is
associated with every active check node and connected once to that
check node and once to the next adjacent active check node on the
right.

Using a similar derivation to that for LDPC codes \cite{Kudekar_IT11}, leads to the rate of the $(q, a, L, w)$ RA ensemble as:
\begin{align}
    \label{equ:ratewra}
    r_{\text{RA},w} &= \frac{2L+1}{2L+1 + \frac{q}{a} \left[ 2L-w + 2\left( w + 1 - \sum_{i=0}^{w}\left(\frac{i}{w}\right)^{a}\right)\right]}.
\end{align}

\subsection{Encoding}

The motivation for considering SC-RA codes is their low encoding
complexity. As for traditional repeat-accumulate codes, SC-RA codes
can be encoded with complexity linear in the code length by the
serial concatenation of a repetition code, interleaver, combiner and
$\frac{1}{1+D}$ convolutional encoder or accumulator.

RA and SC-RA codes are systematic so that the message bits make up
the first $K$ bits in the codeword meaning that codeword bits can be
transmitted as soon as message bits are received. The structure of
SC-RA codes also has the additional advantage of limiting the number
of message bits that must be received before the first parity bit
can be encoded. Consider fig.~\ref{fig:GraphRAq4}.
A parity bit in the $i$th location is a function only of message bits in the $i$th and previous $q-1$ locations.

\section{Density Evolution} \label{sec:DE}

In this section we derive closed form expressions for density
evolution for the $(q,a,L,w)$ ensemble on the BEC and show how the
multi-edge formulation for LDPC codes can be used to derive
thresholds for the $(q,a,L)$ ensemble.

Following a similar approach to that used for the LDPC $w$-ensemble
\cite{Kudekar_IT11}, gives density evolution equations for the SC-RA $(q,a,L,w)$
ensemble:
\begin{subequations}
\label{equ:dewra}
\begin{align}
    x_{i}^{(\ell+1)} &= \epsilon \left(1-\frac{1}{w}\sum_{j=0}^{w-1}\left[\rule{0cm}{7mm}\right.\left(1-y_{i+j}^{(\ell)}\right)^{2}\right.\nonumber\\
    &{}\hspace{16mm}\cdot\left.\left.\left(1-\frac{1}{w}\sum_{k=0}^{w-1}x_{i+j-k}^{(\ell)}\right)^{a-1}\right]\right)^{q-1}\\
    y_{i}^{(\ell+1)} &= \epsilon \left(1-\left(1-y_{i}^{(\ell)}\right)\left(1-\frac{1}{w}\sum_{k=0}^{w-1}x_{i-k}^{(\ell)}\right)^{a}\right),
\end{align}
\end{subequations}
where $x_{i}^{(\ell)}$ and $y_{i}^{(\ell)}$ denote the erasure
probabilities from message bits and parity bits respectively at
position $i$, at iteration $l$.

Density evolution for the $(q,a,L)$ ensembles results in more
complicated expressions since the erasure probabilities on edges
connected to one protograph cannot be averaged as for the
$(q,a,L,w)$ ensemble above. While it is still possible to write the
expressions in closed form we instead choose the multi-edge
framework to represent the structure of the $(q,a,L)$ ensemble and
use multi-edge density evolution to evaluate the decoding thresholds
over the erasure channel. For a detailed description of multi-edge
density evolution we refer the reader to \cite[Sec.
7]{Richardson-2008-B}.

\begin{figure}[t!]
    \begin{center}
            \subfloat[$(q,a,L,w=q)$ and $(d_l,d_r,L,w=d_l)$ ensembles]{\label{fig:de_w}\includegraphics[width=0.99\columnwidth]{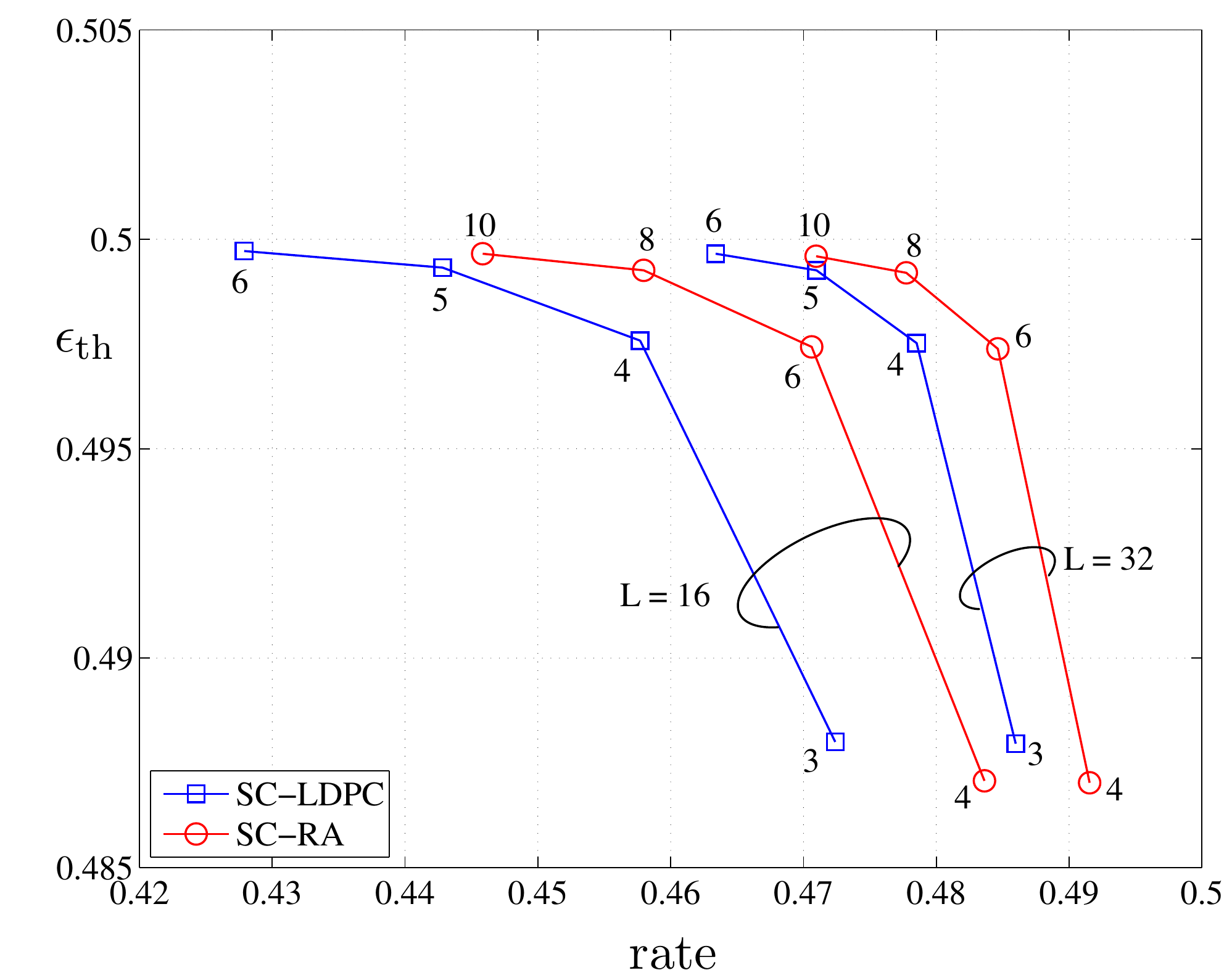}}\\
            \subfloat[$(q,a,L)$ and $(d_l,d_r,L)$ ensembles]{\label{fig:de_met}\includegraphics[width=0.99\columnwidth]{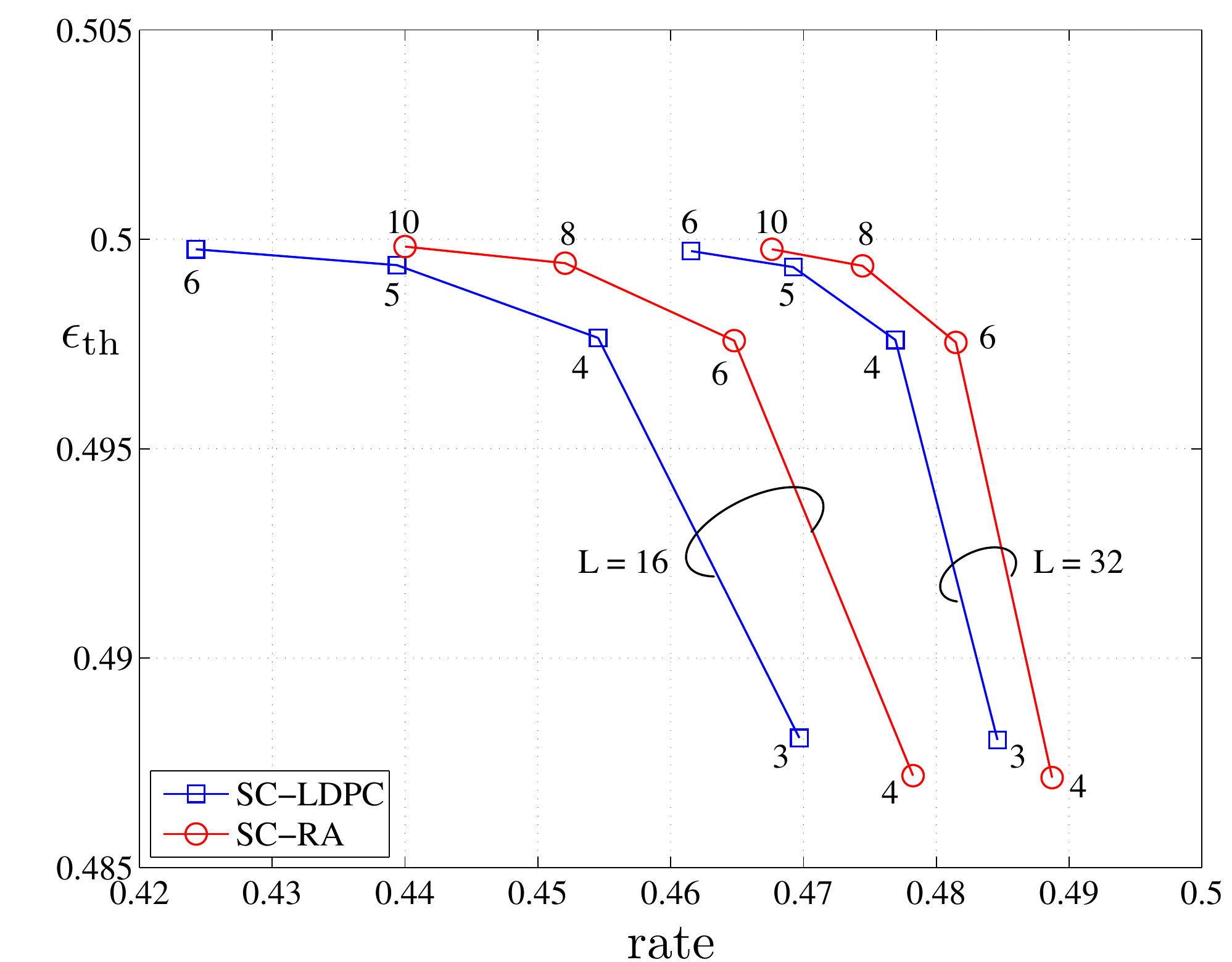}}
        \caption{Thresholds versus rates of spatially coupled LDPC and RA codes. The markers represent the different variable node degrees of the message bits, $d_l = \{3, 4, 5, 6\}$ and $q =
\{4, 6, 8, 10\}$ (lower degrees correspond to lower thresholds).}
            \label{fig:de}
    \end{center}
\end{figure}

Numerical results are shown in fig.~\ref{fig:de} where we compare
the decoding thresholds and rates (\cite[Equ.
7]{Kudekar_IT11},\cite[Lemma 3]{Kudekar_IT11} for the LDPC ensemble
and \eqref{equ:ratewra} and \eqref{equ:dewra} for the RA ensemble).
Each curve corresponds to a value of $L$ and the markers represent
the variable node degree of the message bits. Higher degrees lead to
an improved decoding threshold but result in a lower rate due to the
increasing number of additional check nodes at the ends of the
graph.

The $(q,a,L,w)$ ensembles are shown for the case $w=q$ and so these
ensembles will not have extra check nodes over those in the
$(q,a,L)$ ensembles with the same parameters. Consequently the
$(q,a,L,w=q)$ ensembles have a slightly higher rate than the
$(q,a,L)$ ensembles with the same parameters, due to the likelihood
of some check nodes not being active for a given code. When $w$ is
chosen to be larger than $q$ there is also the likelihood of extra
check nodes, outside of those used in the $(q,a,L)$ ensemble,
becoming active and thereby slightly reducing the code rate.

Note that due to the accumulator (which consists of degree 2
variable nodes), SC-RA codes have a lower average variable node
degree than SC-LDPC codes with the same degree for the message bits.
To compare LDPC and RA codes with the same densities, and hence
similar decoding complexities, we compare an LDPC base code with bit
degree $d_l$ with an RA base code with bit degree $q =
\frac{1}{r}(d_l -2)+2$ where r is the code rate. Thus
fig.~\ref{fig:de} shows points for LDPC protographs with $d_l = \{3,
4, 5, 6\}$ and RA protographs with $q = \{4, 6, 8, 10\}$.

We observe that SC-RA codes perform better than SC-LDPC codes giving
a higher code rate at the same decoding threshold as the SC-LDPC
codes.

\section{Simulation Results} \label{sec:Sims}
In this section we randomly construct SC-RA codes and compare their
decoding performance at finite lengths to SC-LDPC codes. Consider
for example the $(q,a,L)$ ensemble with thresholds shown in
fig.~\ref{fig:de} for $L=16$ with $q=6$ for the SC-RA code and
$d_{l}=4$ for the SC-LDPC code. The SC-RA ensemble has an average
variable node degree of $3.86$ (compared to $4$ for the SC-LDPC
code), a higher rate and a similar decoding threshold.

\begin{figure}[h]
    \begin{center}
        \includegraphics[width=\columnwidth]{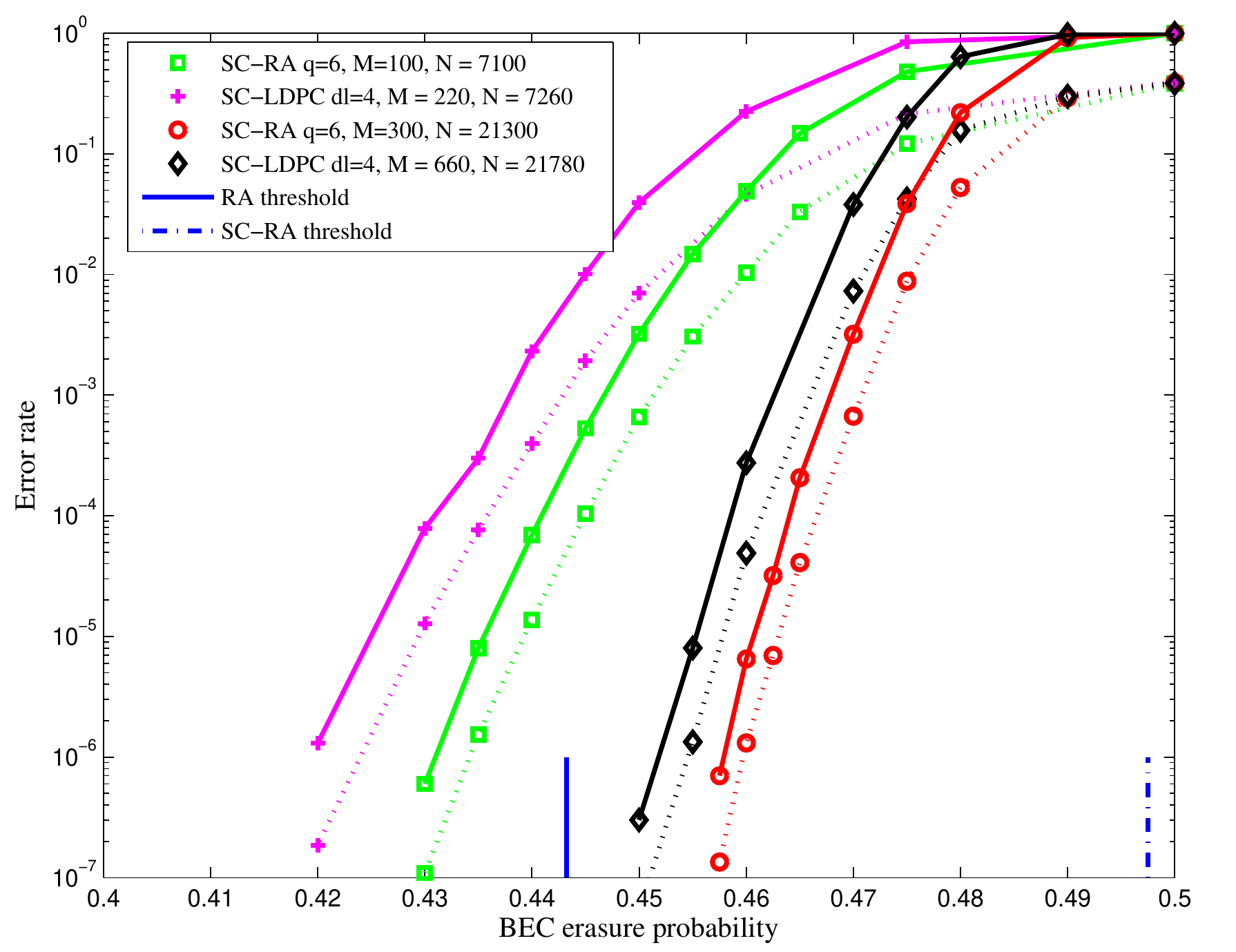}
        \caption{Erasure correction performance of SC-LDPC and SC-RA codes with L=16,
        for $K=3,300$ and $K=9,900$ using iterative decoding with a maximum of 1000 iterations. Solid lines show the word erasure rate and dashed lines show the bit erasure rate.}
        \label{fig:SC-RAvsSC-LDPC_BER}
    \end{center}
\end{figure}

Fig.~\ref{fig:SC-RAvsSC-LDPC_BER} shows the erasure correction
performance of (6,6,16) SC-RA codes with $M$ set to 100 and 300
respectively. Also shown is the performance of (4,8,16) SC-LDPC
codes with $M$ set to 220 and 660 respectively. (Recall that for
SC-LDPC codes $M$ specifies the number of all bit nodes, whereas for
SC-RA codes $M$ specifies the number of message bit nodes). For the
two shorter codes, each code transmits 3,300 message bits, however
the SC-RA code has a slightly higher rate requiring only 7100
codeword bits (r = 0.4648) instead of 7260 (r = 0.4545). For the two
longer codes, each code transmits 9,900 message bits, however the
SC-RA code requires only 21300 codeword bits instead of 21780.

Also shown is the threshold for SC-RA codes (from fig.~\ref{fig:de})
and the iterative decoding threshold for RA codes with the same
degree distribution (and rate) as the SC-RA codes but without the
spatial coupling.

In fig.~\ref{fig:SC-RAvsSC-LDPC_BER} we can see that spatial
coupling or RA codes does indeed produce codes with excellent
iterative decoding performance. We also see that the performance of
the SC-RA codes is better than that of the SC-LDPC codes with
similar decoding complexity (as measured by the Tanner graph edge
density) despite both having the same threshold. We suspect that for
finite length codes the structure of the SC-RA codes gives them a
further advantage (in addition to the slightly higher rate for the
same threshold) over LDPC codes.

\section{Discussion}

In this paper we have proposed a new class of spatially coupled
codes based on repeat-accumulate protographs. We show that spatially
coupled repeat-accumulate codes have several advantages over
spatially coupled low-density parity-check codes including simpler
encoders and slightly better thresholds than spatially coupled
low-density parity-check codes with similar rates and decoding
complexity. Simulation results for finite-length spatially coupled
repeat-accumulate codes also show improved decoding performances
over spatially coupled low-density parity-check codes with the same
threshold.

\bibliographystyle{ieeetran}
\bibliography{SCcodes}

\end{document}